\documentclass[11pt,twoside]{article}

\usepackage{fullpage}
\usepackage[super,sort,comma]{natbib}

\usepackage{multirow}
\usepackage{amssymb}
\usepackage{amsmath}
\usepackage{fancyhdr}		
\usepackage[table]{xcolor} 
\definecolor{pastelyellow}{RGB}{255,255,204}

\usepackage[section]{placeins}   %

\usepackage{graphicx}
\usepackage{booktabs}
\usepackage{subcaption}

\makeatletter \renewcommand\@biblabel[1]{$^{#1}$} \makeatother
\newcommand{\red}[1]{\textcolor{red}{#1}}

\usepackage[mathlines]{lineno}

\usepackage{amsmath}

\usepackage{xcolor}

\definecolor{gray}{rgb}{0.6,0.6,0.6}
\definecolor{red}{rgb}{0.85,0,0}
\definecolor{green}{rgb}{0,0.85,0}
\definecolor{blue}{rgb}{0,0,0.85}
\definecolor{beige}{rgb}{0.92,0.87,0.78}

\newcommand{\method}{VI-PRISM}

\input{command-template}

\begin{document}

\begin{centering}

  {\bf{\LARGE{Perfusion Imaging and Single Material Reconstruction in Polychromatic Photon Counting CT}}}

\vspace*{.2in}

{\large{
\begin{tabular}{ccc}
Namhoon Kim$^{\dagger}$, Ashwin Pananjady$^{\star,\dagger}$, Amir Pourmorteza$^{\ddagger,\S, \P}$, and Sara Fridovich-Keil$^{\dagger}$
\end{tabular}
}}
\vspace*{.2in}

\begin{tabular}{c}
$^{\star}$School of Industrial and Systems Engineering , Georgia Institute of Technology\\
$^{\dagger}$School of Electrical and Computer Engineering, Georgia Institute of Technology\\
$^{\ddagger}$Department of Radiology and Imaging Sciences, Emory University School of Medicine \\
$^{\S}$Winship Cancer Institute, Emory University \\
$^{\P}$Department of Biomedical Engineering, Georgia Institute of Technology\\

\end{tabular}

\begin{tabular}{c}

\end{tabular}
\end{centering}

\vspace*{.2in}

{\centering \today\par}

\vspace*{.2in}

\begin{abstract}

\noindent {\bf Background:} Perfusion computed tomography (CT) images the dynamics of a contrast agent through the body over time, and is one of the highest X-ray dose scans in medical imaging. Recently, a theoretically justified reconstruction algorithm based on a monotone variational inequality (VI) was proposed for single material polychromatic photon-counting CT, and showed promising preliminary results at low-dose imaging.

\noindent {\bf Purpose:} We adapt this reconstruction algorithm for perfusion CT, to reconstruct the concentration map of the contrast agent while the static background tissue is assumed known; we call our method \method{} (VI-based PeRfusion Imaging and Single Material reconstruction). We evaluate its potential for dose-reduced perfusion CT, using a digital phantom with water and iodine of varying concentration.

\noindent {\bf Methods:} Simulated iodine concentrations range from 0.05 to 2.5 mg/ml. The simulated X-ray source emits photons up to 100 keV, with average intensity ranging from $10^5$ down to $10^2$ photons per detector element. The number of tomographic projections was varied from 984 down to 8 to characterize the tradeoff in photon allocation between views and intensity.

\noindent {\bf Results:} We compare \method{} against filtered back-projection (FBP), and find that \method{} recovers iodine concentration with error below 0.4 mg/ml at all source intensity levels tested. Even with a dose reduction between $10\times$ and $100\times$ compared to FBP, \method{} exhibits reconstruction quality on par with FBP.

\noindent {\bf Conclusion:} Across all photon budgets and angular sampling densities tested, \method{} achieved consistently lower RMSE, reduced noise, and higher SNR compared to filtered back-projection. Even in extremely photon-limited and sparsely sampled regimes, \method{} recovered iodine concentrations with errors below 0.4 mg/ml, demonstrating that \method{} can support accurate and dose-efficient perfusion imaging in photon-counting CT.

Code is available at \url{https://github.com/voilalab/VI-PRISM}.

\end{abstract}

\pagenumbering{arabic}
\setcounter{page}{1}

\section{Introduction}

Perfusion CT aims to quantify the kinetics of a contrast agent such as iodine within tissue, providing a quantitative measure of blood flow and volume\cite{zeng2017low}. Accurate estimation of iodine concentration requires separating the contrast signal from the underlying tissue attenuation, often under low-dose conditions where photon statistics are limited\cite{forbrig2024optimizing}.

Photon-counting detector (PCD) CT has finer spatial resolution, more energy bins and improved spectral separation than conventional energy-integrating CT\cite{symons2017photon,iwanczyk2009photon,taguchi2013vision,schlomka2008experimental,leng2019photon,willemink2018photon}, allowing more accurate material-specific imaging\cite{holmes2023ultra, mccollough2023clinical}. These capabilities make PCD-CT a strong platform for quantitative perfusion studies, where the objective is to recover iodine concentration maps that reflect contrast dynamics while minimizing radiation dose.

Recently, an iterative first-order reconstruction algorithm based on a monotone variational inequality (VI) was proposed for single-material PCD-CT \cite{exact}. We adapt this algorithm for perfusion CT, assuming two basis components --- iodine (unknown) and background tissue (known); we call our method \method{}: Variational Inequality based PeRfusion Imaging and Single Material reconstruction.
The algorithm uses a polychromatic forward model and enforces physical constraints such as nonnegativity and spatial regularity. Instead of minimizing an explicit loss function, as is common in iterative and model-based reconstruction algorithms\cite{pourmorteza2016reconstruction, thibault2007three, barber2016algorithm, fridovich2023gradient}, it uses the monotone variational inequality formulation to drive the mismatch between measured and predicted data toward zero while maintaining these constraints. In CT, this provides a provably stable\cite{exact} and physically consistent approach to reconstruct material maps from nonlinear photon-counting data under realistic noise.

Our study in this paper compares \method{} with conventional filtered back-projection (FBP)\cite{herman1976convolution, FDK} across varying photon budgets and sparse projection sampling densities. The goal is to assess whether \method{} can deliver accurate iodine quantification at substantially reduced photon counts --- indeed, such a method would be a good candidate to support dose-efficient and quantitative perfusion imaging.

\section{Methods}

\subsection{Algorithm}

Our reconstruction approach is based on a polychromatic forward model for PCD CT, where the expected number of transmitted photons is given by \cite{schmidt2023constrained}:
\begin{equation}\label{eq:fullforwardmodel}
    \mathbb{E}[y_{\omega,i}(x)]=\overline{I}_{\omega}\sum_{j=1}^{W}s_{\omega,j}\exp\left(-\sum_{m=1}^{M}\sum_{k=1}^{d}\mu_{m,j}A_{i,k}x_{k,m}\right).
\end{equation}
Here, $\omega$ indexes the detector windows, $i$ indexes the measurements (each detector element in each view), and $j$ indexes the $W$ X-ray energies in the source spectrum. 
We use $\overline{I}_{\omega}$ to denote the mean incident photon count in detector window $\omega$, and $s_{\omega,j}$ to denote the normalized detector sensitivity for each detector window and individual X-ray wavelength. 
The unknown image is parameterized by $x_{k,m}$, where $k$ indexes the $d$ spatial locations (vectorized pixels or voxels) and $m$ indexes the $M$ basis materials. 
The mass attenuation coefficient of material $m$ at wavelength $j$ is denoted by $\mu_{m,j}$, and $A_{i,k}$ is the projection matrix element linking pixel $k$ to measurement $i$.

Because PCDs reject electronic noise, we can model the noisy measurements with a Poisson distribution: 
\[ y \sim \mathrm{Poisson}\!\left(\mathbb{E}[y(x)]\right). \] 
Our reconstruction strategy, however, is agnostic to the noise distribution; it relies only on the expectation relation~\eqref{eq:fullforwardmodel} that governs the forward model.

Since our goal is to reconstruct a single material map, we focus on recovering the concentration map of the iodine contrast agent and treat all other material maps as known. 
Accordingly, it is convenient to write the forward model as
\begin{equation}\label{eq:singlematerialforwardmodel}
    \mathbb{E}[y_{\omega,i}(x_\text{iodine})]=\overline{I}_{\omega}\sum_{j=1}^{W}s_{\omega,j}\exp\left(-\sum_{m=1}^{M-1}\sum_{k=1}^{d}\mu_{m,j}A_{i,k}x_{k,m}\right)\exp\left(-\sum_{k=1}^{d}\mu_{\text{iodine},j}A_{i,k}x_{k,\text{iodine}}\right),
\end{equation}
where all variables are known except for the iodine concentration map $x_{\text{iodine}} \in \mathbb{R}^{d}$. This concentration map is to be recovered by our method, \method{}.

\method{} follows an iterative update of the form
\begin{equation}\label{eq:monotoneupdate}
x^{t+1} = P_\mathcal{X}(x^t - \alpha_t F(x^t)),
\end{equation}
where $x^t$ denotes the estimate of the iodine concentration map at iteration $t$, $\alpha_t$ is the step size, $P_\mathcal{X}$ is a projection onto a feasible set $\mathcal{X}$, and $F(x^t): \mathbb{R}^d \rightarrow \mathbb{R}^d$ is a monotone operator defined as \cite{exact}:
\begin{equation}\label{operator}
  F(x) = \frac{1}{3n} \sum_{\omega=1}^3\sum_{i=1}^{n} \big[y_{\omega,i} - \mathbb{E}[y_{\omega,i}(x)] \big] \cdot a_{i}.
\end{equation}
Here $a_i \in \mathbb{R}^d$ is the vector whose entries are $A_{i,k}$ for all pixels $k$, $y_{\omega,i}$ is a measurement corrupted by Poisson noise, and $\mathbb{E}[y_{\omega,i}(x)$ is the noiseless forward model of \Cref{eq:singlematerialforwardmodel} evaluated at the current iodine concentration map estimate $x = x^t$.
The operator $F$ is not the gradient of a loss function, but it plays a similar role algorithmically. In particular, it has a \emph{monotonicity} property that ensures convergence of the iteration~\eqref{eq:monotoneupdate} to its fixed point provided the step sizes are chosen appropriately~\cite{exact}. Additionally, this fixed point is expected to be accurate since by construction, the true iodine map $x^\star$ is a fixed point of the iteration~\eqref{eq:monotoneupdate} in expectation, i.e., we have $\mathbb{E}[F(x^\star)] = 0$.

The projection step $P_\mathcal{X}$ constrains each iterate $x^t$ to lie in a feasible set $\mathcal{X}$ that captures prior knowledge about the iodine map. In our experiments, $\mathcal{X}$ is the intersection of the nonnegative orthant (generalization of a 2D quadrant to $n$-dimensional space), since concentrations must be nonnegative, and a total variation (TV) ball \cite{candesrombergtao}, which encourages the reconstructed iodine map to be piecewise constant.

The iterations proceed until convergence, which we define based on the running average of the iterates stabilizing within a small threshold.

Our \method{} algorithm~\eqref{eq:monotoneupdate} is based upon the EXACT (Extragradient Algorithm for CT) method recently proposed by Lou et al. \cite{exact}, with two key modifications. Both methods assume a polychromatic X-ray source, both methods reconstruct the concentration map of a single material, and both methods leverage the same monotone operator $F$ (\Cref{operator}). However, the iteration rule in \Cref{eq:monotoneupdate} is simpler than the update rule of EXACT, while inheriting the same desirable theoretical guarantees of convergence and noise-robust image recovery (for certain step sizes) \cite{exact}. We adopt the simpler update rule in \Cref{eq:monotoneupdate} for \method{} because it empirically provides faster convergence than the extragradient update in EXACT (though the convergence rate for EXACT is faster in theory). 

The other key distinction between \method{} and EXACT \cite{exact} is that EXACT assumes the imaging target consists of a single material, while \method{} allows the target to contain an arbitrary combination of materials as long as only one material map is unknown. This modification is reflected in \Cref{eq:singlematerialforwardmodel}, in which we incorporate an arbitrary set of known material maps while reconstructing a single unknown material map. \method{} is thus applicable to perfusion imaging, in which a static pre-scan of the patient at higher X-ray dose can provide accurate material maps for all materials except the contrast agent. The concentration map of the contrast agent can then be reconstructed over time by \method{}, ideally at much lower X-ray dose.

\subsection{Phantom}
\begin{figure}[htbp]
    \centering
    \includegraphics[width = \textwidth]{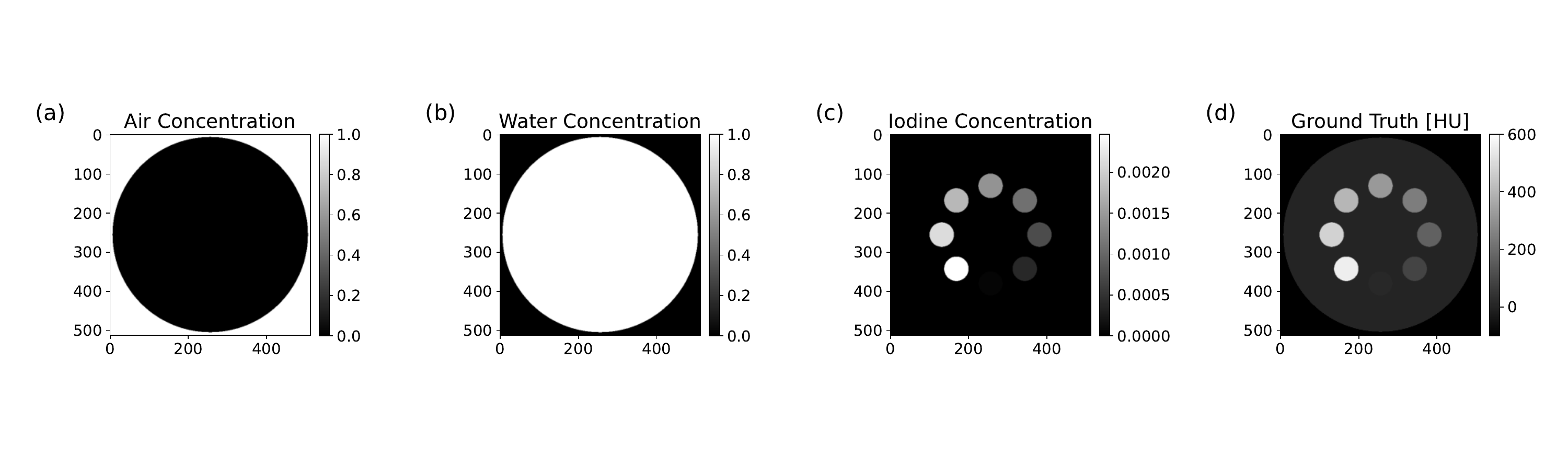}
    \caption{Visualization of the spatial distribution of (a) air, (b) water, and (c) iodine in the ground truth phantom, along with (d) the ground truth HU image. 
    }
    \label{fig:phantom}
\end{figure}

We use a digital phantom based on the cylindrical ACR CT phantom\cite{mccollough2004phantom}, with diameter 200~mm. The background materials are air and water, and we add eight circular ROIs with distinct concentrations of iodine diluted in water to assess quantitative performance in a perfusion task with iodine as the contrast agent. Let $x \in \mathbb{R}^{513 \times 513}$ denote the unknown iodine concentration map, with $d=513^2$ pixels arranged in a square grid. For our algorithm, the water and air distributions are treated as known and fixed, and iodine is the only unknown material to be reconstructed, while FBP knows nothing about the materials. In practice, the static concentration maps of all materials present before introduction of contrast agent could be obtained by a higher-dose pre-scan, so that only the dynamic concentration map of the contrast agent is unknown during low-dose perfusion CT. The ROI layout follows a circular arrangement, with eight positions approximately aligned with clock angles at 12, 1:30, 3, 4:30, 6, 7:30, 9, and 10:30. Iodine concentrations in the eight inserts are constant within each ROI and different across ROIs (0.05, 0.39, 0.74, 1.09, 1.43, 1.78, 2.12, 2.47~mg/ml). The phantom is illustrated in \Cref{fig:phantom}.

\subsection{CT scanner simulation setup}

We simulate a fan-beam CT system using TIGRE \cite{tigre} with source–object and source–detector distances of $\mathrm{DSO}=625.61~\mathrm{mm}$ and $\mathrm{DSD}=1097.6~\mathrm{mm}$, respectively. The detector consists of $\nu=2N_x=1026$ channels with a pitch of approximately $1.09~\mathrm{mm}$. The field of view is $220 \times 220~\mathrm{mm}^2$, and reconstruction is performed on a $513 \times 513$ grid (voxel size $\approx 0.43 \times 0.43~\mathrm{mm}^2$). The incident X-ray beam was modeled using a diagnostic spectrum from Bhat et al. \cite{bhat1998diagnostic}, covering photon energies up to 100 keV with a spectrum discretization interval of 2 keV. The spectrum was log-interpolated to this discrete grid of energies, and normalized so that the total probability sums to one. Detector sensitivities were modeled as smooth spectral response functions applied across the same discrete set of energies, with three overlapping detector energy bins following Barber and Sidky \cite{barber2024convergence}, at energy ranges of approximately 5--55 keV, 45--75 keV, and 65--100 keV. Each response function is constructed by multiplying the incident spectrum with a smooth, blurred weighting profile in energy, ensuring finite resolution and overlapping sensitivity across neighboring energies. At every energy, these response curves are constant across all rays and are scaled so that their sum exactly reproduces the incident source spectrum.

\subsection{Reconstruction}

We compare our \method{} against FBP as a baseline. \method{} takes raw noisy projection photon counts as input, while for FBP we perform noise clipping and logarithmic preprocessing.
FBP uses a Hann filter and the same fan-beam geometry as in data generation, with reconstructions performed separately at each energy and then summed with the source spectrum as weights. 
For \method{} we employ total variation (TV) regularization\cite{candesrombergtao} via projections at each step. Concretely, we project each iterate onto a feasible set $\mathcal{X}$ (see~\eqref{eq:monotoneupdate}), which is given by the intersection of the positive orthant and a TV ball of a certain radius. We set the radius of the TV ball to be equal to the TV of the ground truth iodine map. 

After reconstruction, conversion to Hounsfield Units (HU) is performed differently for the two methods. 
For \method{}, we compute the energy-dependent linear attenuation coefficients $\mu(E)$ and form a weighted average using the source spectrum $s(E)$.

For FBP, we perform calibration-based correction using the reconstructed values in regions of interest corresponding to water and air.

\subsection{Experiment design}

We study the allocation of a fixed total photon budget $B$ across a varying number of views $n_s$. We sweep
\[
n_s \in \{984,\,492,\,246,\,164,\,123,\,82,\,41,\,24,\,12,\,8\},
\]
and four total photon budgets
\[
B \in \{984\times 10^{2},~984\times 10^{3},~984\times 10^{4},~984\times 10^{5}\}.
\]
For each $(B, n_s)$, the average number of photons emitted per detector element is
\[
I = \frac{B}{n_s}.
\]
Projection angles are uniformly spaced over $[0,2\pi)$: $\theta_k = 2\pi k / n_s,\; k=0,\dots,n_s-1$. To assess variability due to Poisson noise sampling, we repeat each simulation over 9 random seeds.

\begin{figure}[htbp]
    \centering
    \includegraphics[width = 0.8\textwidth]{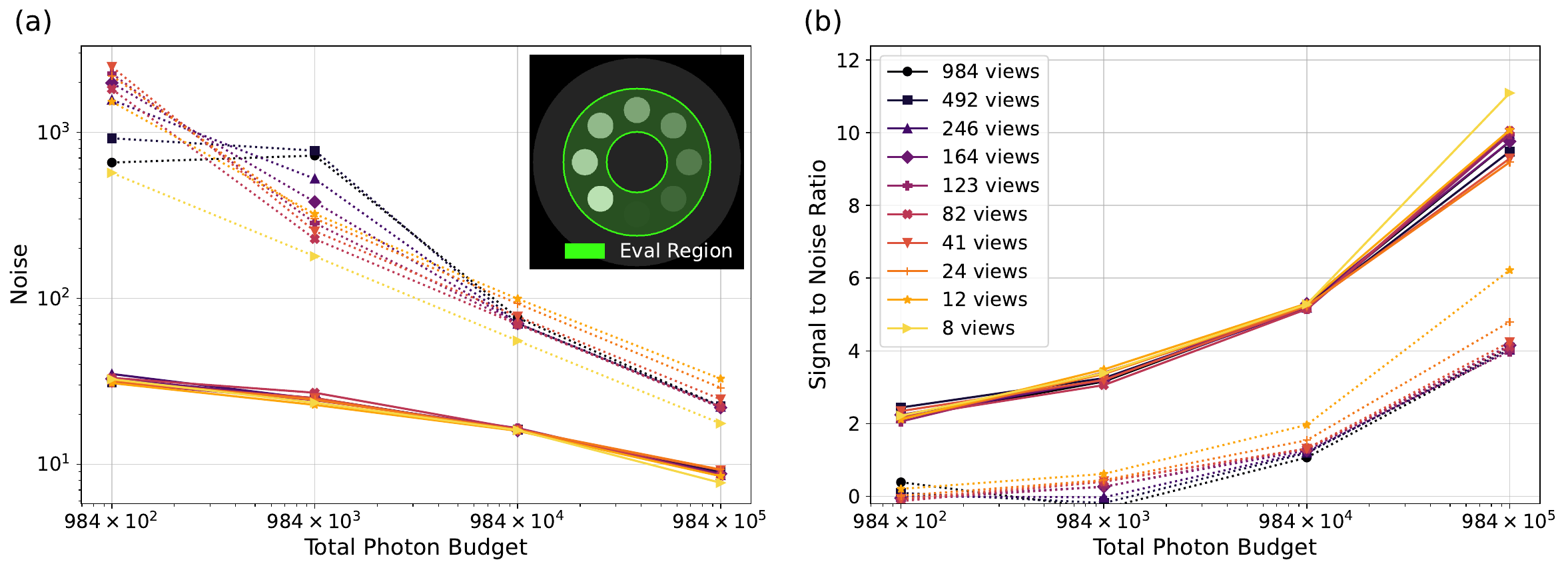}
    \caption{Noise and signal to noise ratio (SNR) as a function of total photon budget and number of projections.
Each curve corresponds to a different number of projections for the given total budget. Solid lines correspond to \method{}, and dotted lines correspond to FBP. Subfigure (a) reports noise and subfigure (b) reports SNR, both evaluated within the ring mask shown in green in the inset.}
    \label{fig:Noise_SNR_vs_total_photon_budgets}
\end{figure}

\begin{figure}[htbp]
    \centering
\includegraphics[width = 0.95\textwidth]{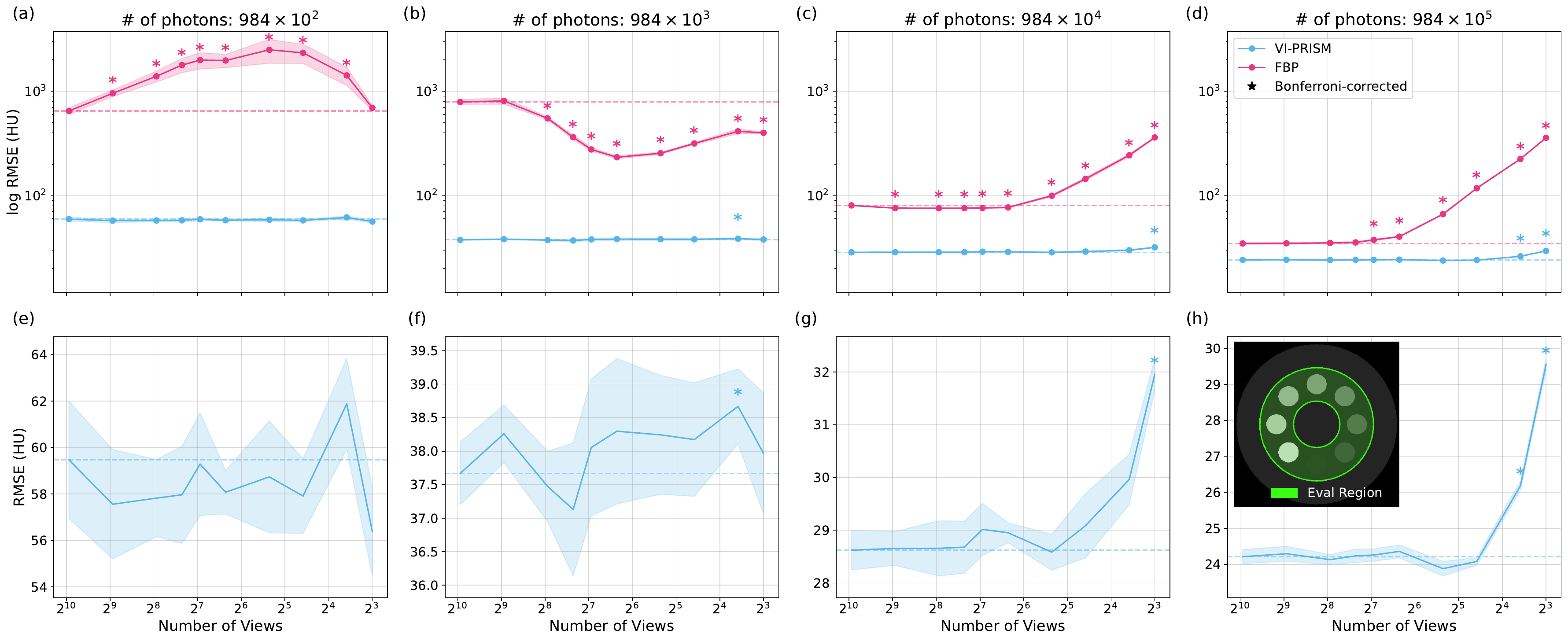}
\caption{RMSE versus number of views, evaluated within the ring mask inset shown in subfigure (h), for both \method{} and FBP under different total photon budgets. 
The first row shows RMSE values for both \method{} and FBP reconstructions, while the second row focuses on \method{}. 
Error bars indicate the standard deviation across 9 random seeds. 
Star annotations indicate statistical significance after Bonferroni correction, denoting whether the RMSE with a smaller number of projections is statistically different from the RMSE with the same photon budget spread among the full set of $984$ projections.}
\label{fig:rmse}
\end{figure}

\begin{figure}[htbp]
    \centering
\includegraphics[width = 0.95\textwidth]{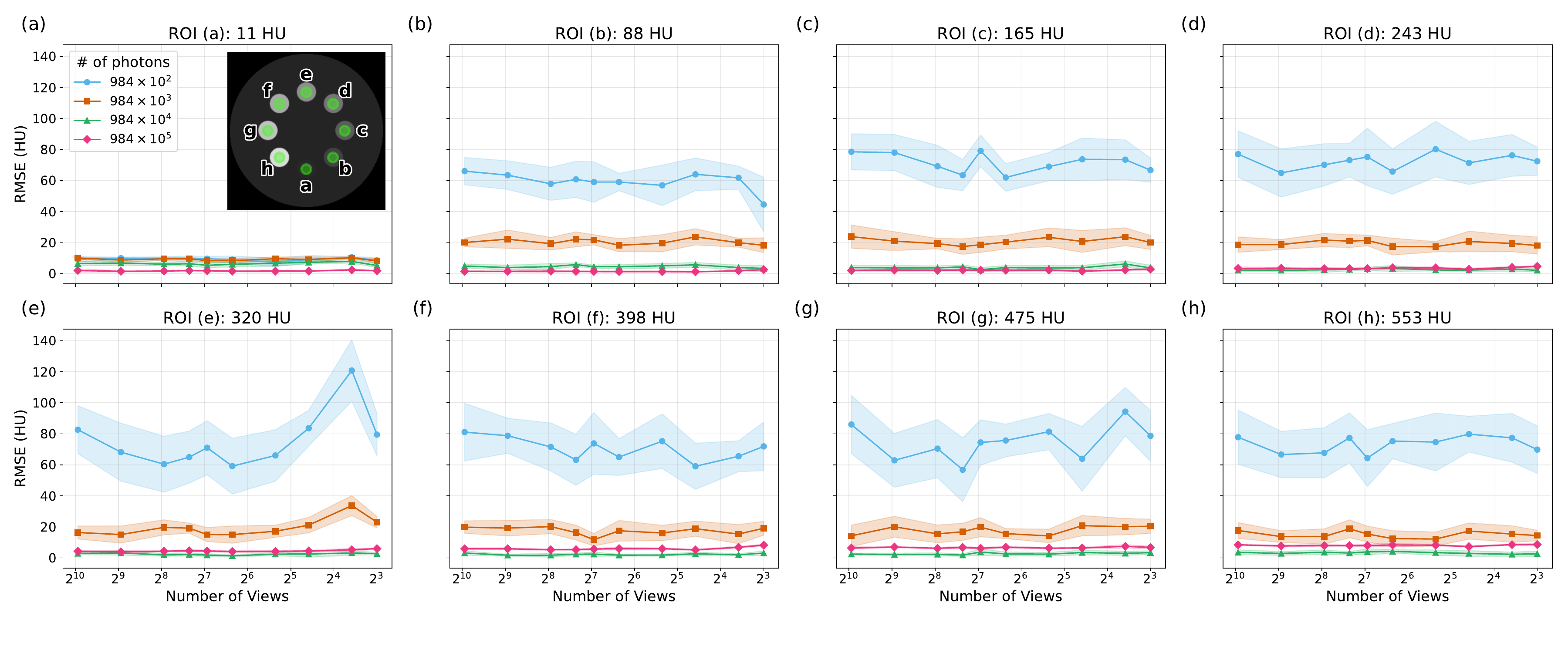}
\caption{HU error for \method{} reconstruction as a function of number of views, for each total photon budget. Error bars indicate the standard deviation across 9 random seeds. Evaluation is within each iodine insert, as visualized in (a); the title of each subfigure details the ground truth CT number (HU) in the corresponding insert.}

\label{fig:HU_errors}
\end{figure}

\begin{figure}[htbp]
    \centering
\includegraphics[width = 0.95\textwidth]{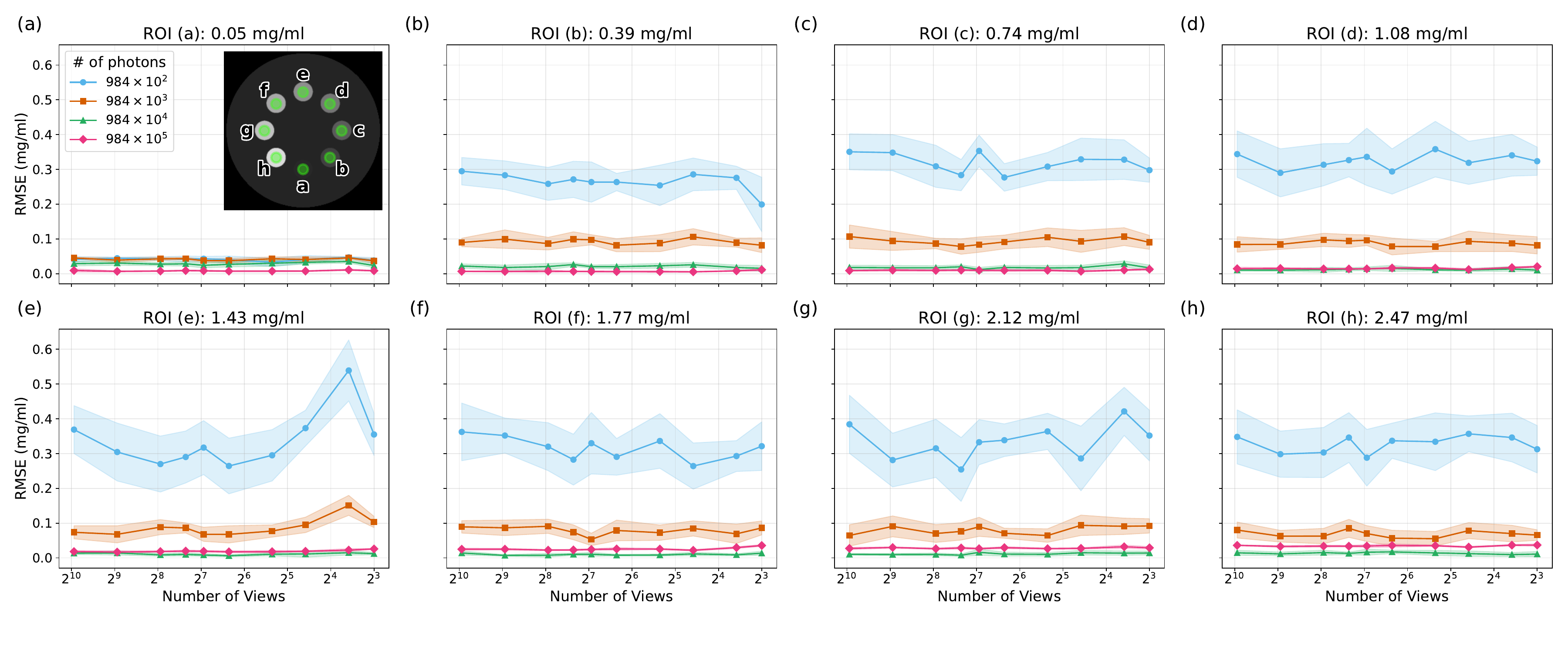}
\caption{Iodine concentration error (mg/ml) for \method{} reconstruction as a function of number of views, for each total photon budget. Error bars indicate the standard deviation across 9 random seeds. Evaluation is within each iodine insert, as visualized in (a); the title of each subfigure details the ground truth iodine concentration in the corresponding insert.}

\label{fig:concentration_errors}
\end{figure}

\begin{figure}[htbp]
    \centering
    \includegraphics[width = \textwidth]{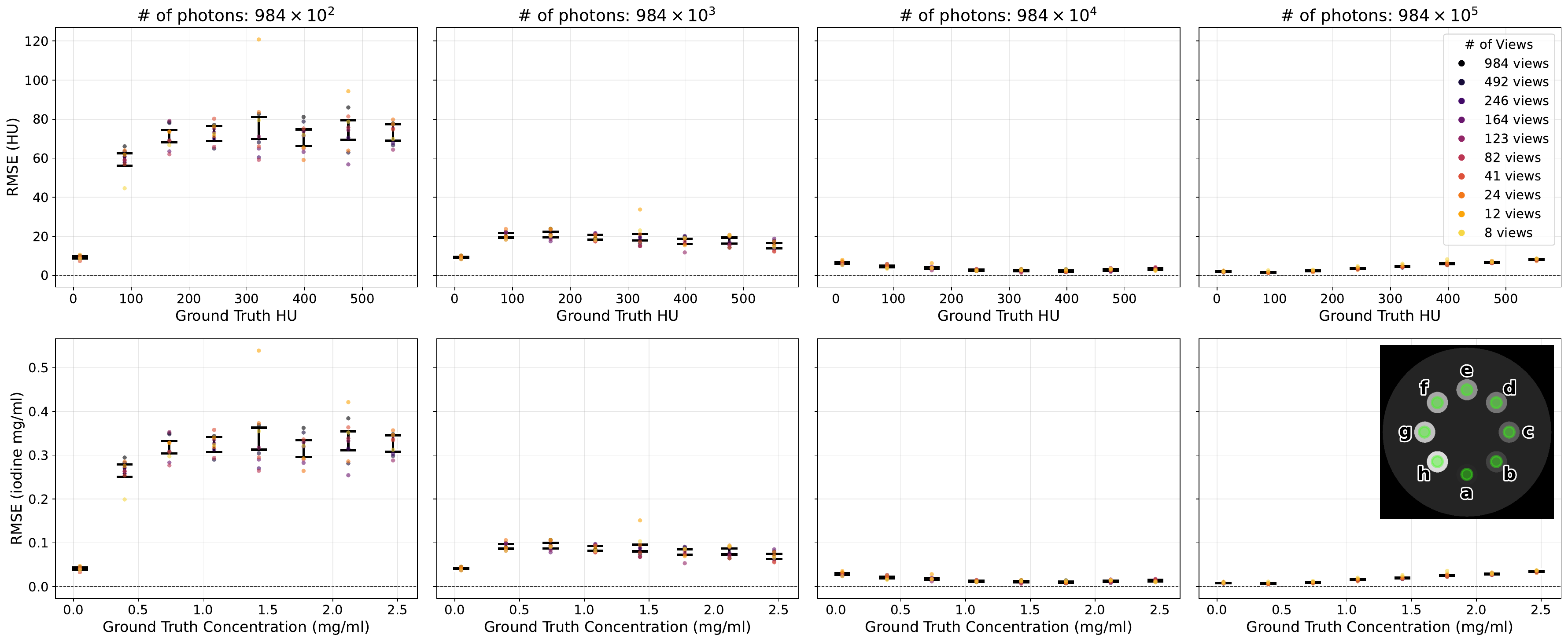}
    \caption{Bland-Altman style plots reporting RMSE in HU (top row) and iodine concentration (bottom row) versus their ground truth values, for \method{} reconstruction with each total photon budget. Error bars indicate 95\% confidence intervals for each iodine-containing ROI, over all simulations with the same total photon budget.}
    \label{fig:blandaltman}
\end{figure}

\begin{figure}[htbp]
    \centering
    \includegraphics[width = 0.95\textwidth]{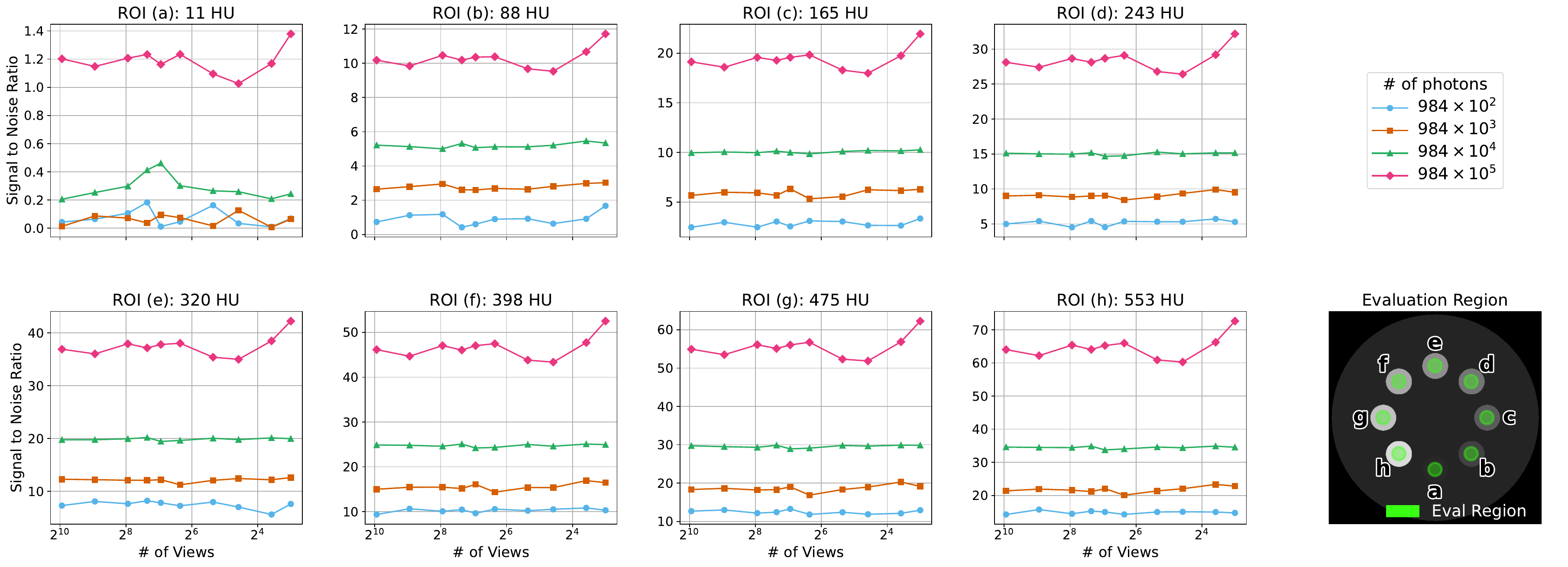}
    \caption{SNR in each iodine-containing ROI as a function of number of projections, for \method{} reconstruction with each total photon budget.}
    \label{fig:snr_vs_roi}
    \end{figure}

Quantitative metrics are evaluated using the eight circular ROI inserts in \Cref{fig:rmse,fig:Noise_SNR_vs_total_photon_budgets}, or a predefined ring shaped mask. 
For ROI-based metrics, only pixels inside each of the eight ROI inserts are considered in \Cref{fig:concentration_errors,fig:HU_errors,fig:blandaltman,fig:snr_vs_roi}.
Each ROI insert has a radius of approximately $12.5~\mathrm{mm}$. 
However, the actual evaluation region is defined as a smaller concentric circle of radius $6.9~\mathrm{mm}$ within each ROI to exclude potential boundary effects.
For global metrics, we use a ring shaped mask centered at the phantom center to capture all ROIs as well as some of the surrounding water, to incorporate geometric fidelity.
The inner and outer radii of the ring are $30.9~\mathrm{mm}$ and $75.6~\mathrm{mm}$, respectively.

\subsection{Quantitative image analysis}

We evaluate reconstruction quality using four complementary metrics: Root Mean Square Error (RMSE), iodine concentration accuracy, image noise, and Signal-to-Noise Ratio (SNR). 
RMSE quantifies the accuracy of reconstructed CT numbers (Hounsfield Units, HU) relative to the ground truth phantom, calculated within the ring shaped mask. For material decomposition, we assess iodine concentration accuracy (mg/ml) inside each of the eight circular inserts. This metric is only reported for \method{}, since FBP cannot recover material-specific concentrations. Image noise is defined as the standard deviation of a difference image formed by subtracting two reconstructions generated from independent noise realizations, normalized by $\sqrt{2}$, and evaluated within the ring shaped mask. Finally, the SNR is defined as the ratio between the mean signal within the ring mask and the noise metric described above.

\subsection{Statistical analysis}
Figures and tables report mean and one standard deviation.
Statistical significance is marked by stars in \Cref{fig:rmse}. Data were first tested for normality using a Shapiro-Wilk test; statistical significance was then evaluated using a Bonferroni-corrected paired t-test, or a Wilcoxon signed-rank test (for any cases that were deemed non-Gaussian), to determine whether the RMSE obtained from fewer than $984$ projections differs from the RMSE will $984$ projections, given a fixed total photon budget.

\section{Results}

\subsection{Comparing FBP and \method{} reconstructions}

\begin{figure}[htbp]
    \centering
    \includegraphics[width = \textwidth]{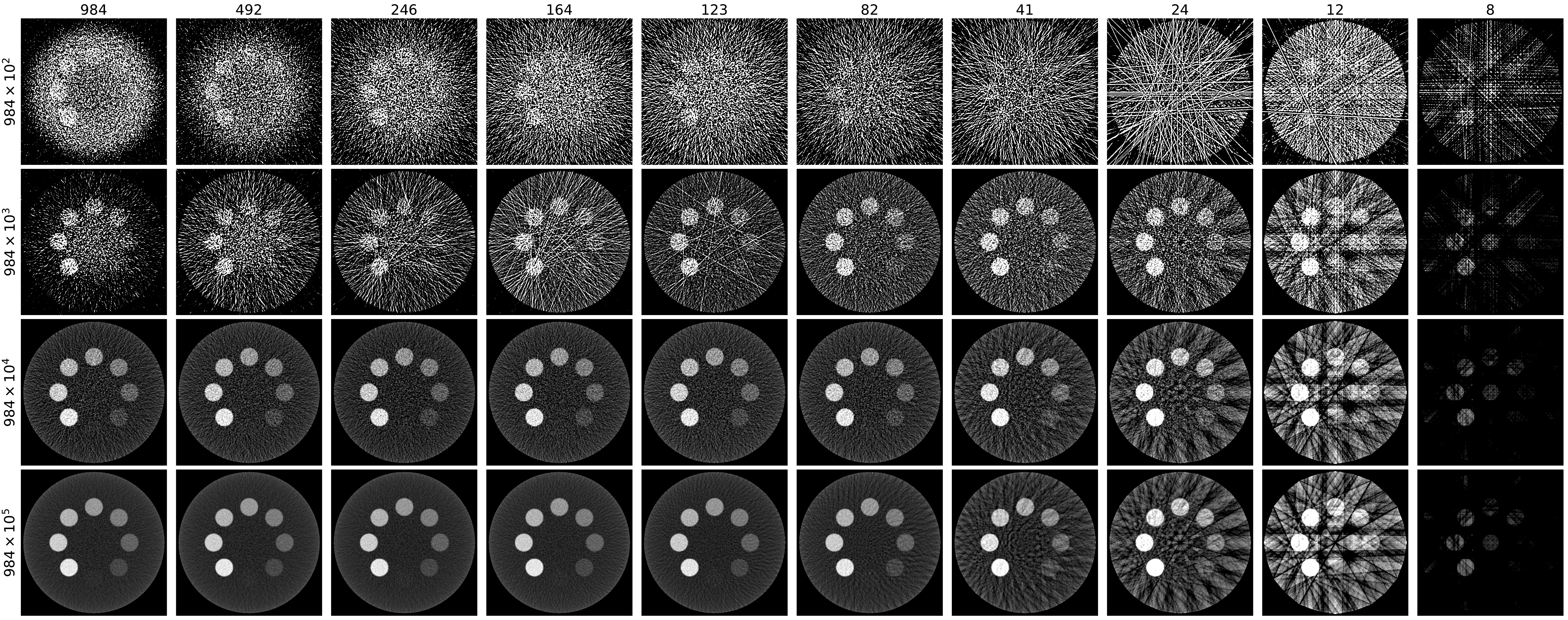}
    \caption{FBP reconstruction results. The photon budget increases tenfold in each row, from $984 \times 10^2$ in the top row to $984 \times 10^5$ in the bottom row. Columns correspond to the number of projections, among which each photon budget is evenly allocated. The grayscale range is clipped to [-100, 600] HU for illustration.}
    \label{fig:fdk_reconstruction}
\end{figure}

\begin{figure}[htbp]
    \centering
    \includegraphics[width = \textwidth]{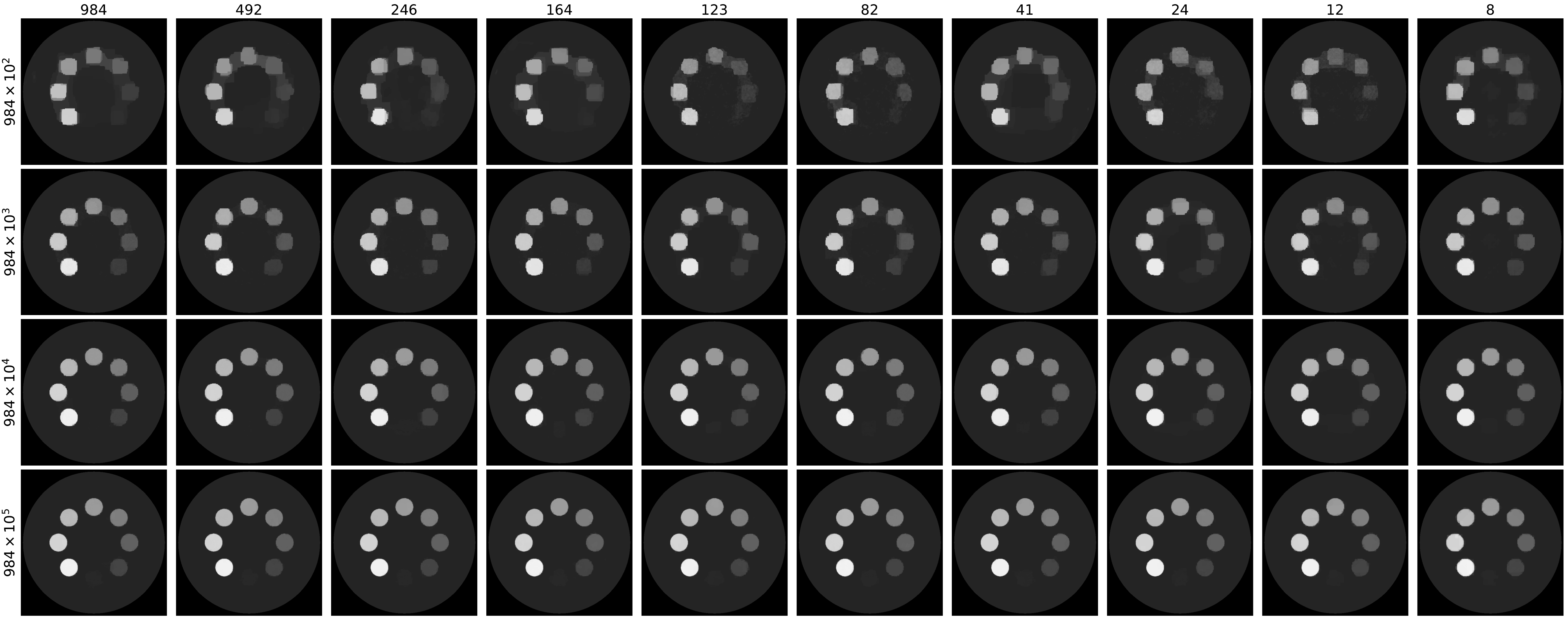}
    \caption{\method{} reconstruction results. The photon budget increases tenfold in each row, from $984 \times 10^2$ in the top row to $984 \times 10^5$ in the bottom row. Columns correspond to the number of projections, among which each photon budget is evenly allocated. The grayscale range is clipped to [-100, 600] HU for illustration.}
    \label{fig:monotone_reconstruction}
\end{figure}

Visual reconstructions across imaging setups are shown in \Cref{fig:fdk_reconstruction} and \Cref{fig:monotone_reconstruction}, for FBP and \method{}, respectively. When the total photon budget is below $984 \times 10^4$ or the number of projections is extremely limited (i.e., fewer than 41 views), FBP reconstructions exhibit severe artifacts and loss of structural fidelity. In contrast, \method{} reconstructions remain robust to angular undersampling down to a total photon budget of $984 \times 10^3$. 

Noise characteristics for both reconstruction methods are summarized in \Cref{fig:Noise_SNR_vs_total_photon_budgets}, showing that \method{} reconstructions consistently achieve lower noise and higher SNR compared to FBP reconstructions, across all photon budgets and numbers of projections. Both algorithms enjoy lower noise and higher SNR with increasing total photon budget, as expected.

\Cref{fig:rmse} visualizes RMSE (in HU) for both reconstruction methods, stratified by both total photon budget and number of projections. We mark with an asterisk any setting for which the RMSE with fewer than $984$ projections differed significantly from the RMSE with $984$ projections and the same total photon budget; significance testing followed a Bonferroni-corrected paired t-test with significance level $p < 0.01$. \method{} reconstructions are uniformly lower-RMSE than FBP reconstructions. The RMSE of the FBP reconstruction with a photon budget of $984 \times 10^5$ is between the RMSE values of the \method{} reconstructions with $984 \times 10^3$ and $984 \times 10^4$ total photons.

Numeric values for the RMSE (in HU) of both algorithms, stratified by total photon budget and number of projections, are summarized in \Cref{tab:hu_views_budgets} and \Cref{tab:hu_views_budgets_fdk}. Since \method{} can also estimate quantitative iodine concentration, \Cref{tab:conc_views_budgets} summarizes the RMSE of its reconstructions in terms of mg/ml of iodine. All tables report results evaluated over the ring mask that captures both the eight iodine-containing inserts and some of the surrounding water.

\begin{table*}[htbp]
\centering
\scriptsize
\setlength{\tabcolsep}{6pt}
\renewcommand{\arraystretch}{1.3}
\caption{HU error (mean $\pm$ standard deviation) for \method{} reconstruction, stratified by number of projections (rows) and total photon budget (columns). 
Evaluation was performed inside the ring mask visualized in \Cref{fig:rmse}.}
\label{tab:hu_views_budgets}
\begin{tabular}{c|cccc}
\hline
\multirow{2}{*}{\# of views} & \multicolumn{4}{c}{\# of photons} \\
\cline{2-5}
 & $984 \times 10^2$ & $984 \times 10^3$ & $984 \times 10^4$ & $984 \times 10^5$ \\
\hline
984 & 59.5 $\pm$ 2.5 & 37.7 $\pm$ 0.5 & 28.6 $\pm$ 0.4 & 24.2 $\pm$ 0.2 \\
492 & 57.6 $\pm$ 2.4 & 38.3 $\pm$ 0.4 & 28.7 $\pm$ 0.3 & 24.3 $\pm$ 0.2 \\
246 & 57.8 $\pm$ 1.7 & 37.5 $\pm$ 0.5 & 28.7 $\pm$ 0.5 & 24.1 $\pm$ 0.1 \\
164 & 58.0 $\pm$ 2.1 & 37.1 $\pm$ 1.0 & 28.7 $\pm$ 0.5 & 24.2 $\pm$ 0.2 \\
123 & 59.3 $\pm$ 2.2 & 38.1 $\pm$ 1.0 & 29.0 $\pm$ 0.5 & 24.3 $\pm$ 0.2 \\
82 & 58.1 $\pm$ 0.9 & 38.3 $\pm$ 1.1 & 29.0 $\pm$ 0.2 & 24.4 $\pm$ 0.2 \\
41 & 58.7 $\pm$ 2.4 & 38.2 $\pm$ 0.9 & 28.6 $\pm$ 0.3 & 23.9 $\pm$ 0.2 \\
24 & 57.9 $\pm$ 1.6 & 38.2 $\pm$ 0.8 & 29.1 $\pm$ 0.6 & 24.1 $\pm$ 0.1 \\
12 & 61.9 $\pm$ 2.0 & 38.7 $\pm$ 0.6 & 30.0 $\pm$ 0.5 & 26.2 $\pm$ 0.1 \\
8 & 56.4 $\pm$ 1.9 & 38.0 $\pm$ 0.9 & 32.0 $\pm$ 0.3 & 29.5 $\pm$ 0.2 \\
\hline
\end{tabular}
\end{table*}
\begin{table*}[htbp]
\centering
\scriptsize
\setlength{\tabcolsep}{6pt}
\renewcommand{\arraystretch}{1.3}
\caption{HU error (mean $\pm$ standard deviation) for FBP reconstruction, stratified by number of projections (rows) and total photon budget (columns). 
Evaluation was performed inside the ring mask visualized in \Cref{fig:rmse}.}
\label{tab:hu_views_budgets_fdk}
\begin{tabular}{c|cccc}
\hline
\multirow{2}{*}{\# of views} & \multicolumn{4}{c}{\# of photons} \\
\cline{2-5}
 & $984 \times 10^2$ & $984 \times 10^3$ & $984 \times 10^4$ & $984 \times 10^5$ \\
\hline
984 & 647.8 $\pm$ 48.9 & 790.4 $\pm$ 41.5 & 80.6 $\pm$ 0.9 & 34.8 $\pm$ 0.6 \\
492 & 956.0 $\pm$ 67.7 & 808.0 $\pm$ 56.6 & 75.9 $\pm$ 0.6 & 34.9 $\pm$ 0.6 \\
246 & 1390.9 $\pm$ 167.6 & 550.1 $\pm$ 17.5 & 75.7 $\pm$ 0.9 & 35.2 $\pm$ 0.6 \\
164 & 1782.2 $\pm$ 274.4 & 362.6 $\pm$ 17.8 & 75.9 $\pm$ 1.0 & 35.6 $\pm$ 0.6 \\
123 & 1989.0 $\pm$ 356.6 & 276.9 $\pm$ 10.8 & 76.1 $\pm$ 1.0 & 37.6 $\pm$ 0.7 \\
82 & 1971.6 $\pm$ 287.2 & 233.2 $\pm$ 7.4 & 77.0 $\pm$ 1.1 & 40.4 $\pm$ 0.6 \\
41 & 2497.7 $\pm$ 642.0 & 254.6 $\pm$ 6.4 & 99.7 $\pm$ 2.4 & 66.4 $\pm$ 0.7 \\
24 & 2335.3 $\pm$ 488.3 & 316.0 $\pm$ 7.6 & 144.8 $\pm$ 4.5 & 117.7 $\pm$ 1.3 \\
12 & 1418.1 $\pm$ 282.4 & 413.5 $\pm$ 22.6 & 243.5 $\pm$ 7.9 & 224.7 $\pm$ 1.6 \\
8 & 693.7 $\pm$ 39.4 & 399.8 $\pm$ 8.3 & 360.9 $\pm$ 3.8 & 358.3 $\pm$ 1.1 \\
\hline
\end{tabular}
\end{table*}
\begin{table*}[htbp]
\centering
\scriptsize
\setlength{\tabcolsep}{6pt}
\renewcommand{\arraystretch}{1.3}
\caption{Iodine concentration error (mean ± standard deviation) for \method{} reconstruction, stratified by number of projections (rows) and total photon budget (columns).  Evaluation was performed inside the ring mask visualized in \Cref{fig:rmse}.
All of the simulated \method{} reconstructions result in iodine concentration errors within $\pm 0.5$ mg/ml.}
\label{tab:conc_views_budgets}
\begin{tabular}{c|cccc}
\hline
\multirow{2}{*}{\# of views} & \multicolumn{4}{c}{\# of photons} \\
\cline{2-5}
 & $984 \times 10^2$ & $984 \times 10^3$ & $984 \times 10^4$ & $984 \times 10^5$ \\
\hline
984 &  0.265 $\pm$ 0.011 &  0.168 $\pm$ 0.002 &  0.127 $\pm$ 0.002 &  0.108 $\pm$ 0.001 \\
492 &  0.257 $\pm$ 0.011 &  0.170 $\pm$ 0.002 &  0.127 $\pm$ 0.001 &  0.108 $\pm$ 0.001 \\
246 &  0.258 $\pm$ 0.007 &  0.167 $\pm$ 0.002 &  0.127 $\pm$ 0.002 &  0.107 $\pm$ 0.001 \\
164 &  0.258 $\pm$ 0.009 &  0.165 $\pm$ 0.004 &  0.128 $\pm$ 0.002 &  0.108 $\pm$ 0.001 \\
123 &  0.264 $\pm$ 0.010 &  0.170 $\pm$ 0.005 &  0.129 $\pm$ 0.002 &  0.108 $\pm$ 0.001 \\
82 &  0.259 $\pm$ 0.004 &  0.171 $\pm$ 0.005 &  0.129 $\pm$ 0.001 &  0.108 $\pm$ 0.001 \\
41 &  0.262 $\pm$ 0.011 &  0.170 $\pm$ 0.004 &  0.127 $\pm$ 0.002 &  0.106 $\pm$ 0.001 \\
24 &  0.258 $\pm$ 0.007 &  0.170 $\pm$ 0.004 &  0.129 $\pm$ 0.003 &  0.107 $\pm$ 0.000 \\
12 &  0.276 $\pm$ 0.009 &  0.172 $\pm$ 0.002 &  0.133 $\pm$ 0.002 &  0.116 $\pm$ 0.001 \\
8 &  0.251 $\pm$ 0.008 &  0.169 $\pm$ 0.004 &  0.142 $\pm$ 0.001 &  0.131 $\pm$ 0.001 \\
\hline
\end{tabular}
\end{table*}

\subsection{Comparing \method{} reconstructions}

Given the qualitatively and quantitatively superior performance of \method{} compared to FBP, as demonstrated in 

\Cref{fig:fdk_reconstruction,fig:monotone_reconstruction,fig:Noise_SNR_vs_total_photon_budgets,fig:rmse}, 
we next examine the quantitative performance of \method{} in greater detail. \Cref{fig:HU_errors} reports the error of the \method{} reconstruction within each iodine-containing ROI, as a function of the number of projections and with varying total photon budgets. \Cref{fig:concentration_errors} reports these errors in terms of iodine concentration. 
\Cref{fig:blandaltman} summarizes these results in a Bland-Altman style plot, visualizing error in HU and concentration versus their ground truth values. 
\Cref{fig:snr_vs_roi} provides a more detailed view of \Cref{fig:Noise_SNR_vs_total_photon_budgets}, with SNR of the \method{} reconstructions evaluated within each iodine-containing ROI. As expected, SNR increases not only with increasing photon budget but also with increasing iodine concentration.

\subsection{Computational characteristics}

\begin{figure}[htbp]
    \centering
    \includegraphics[width = 0.95\textwidth]{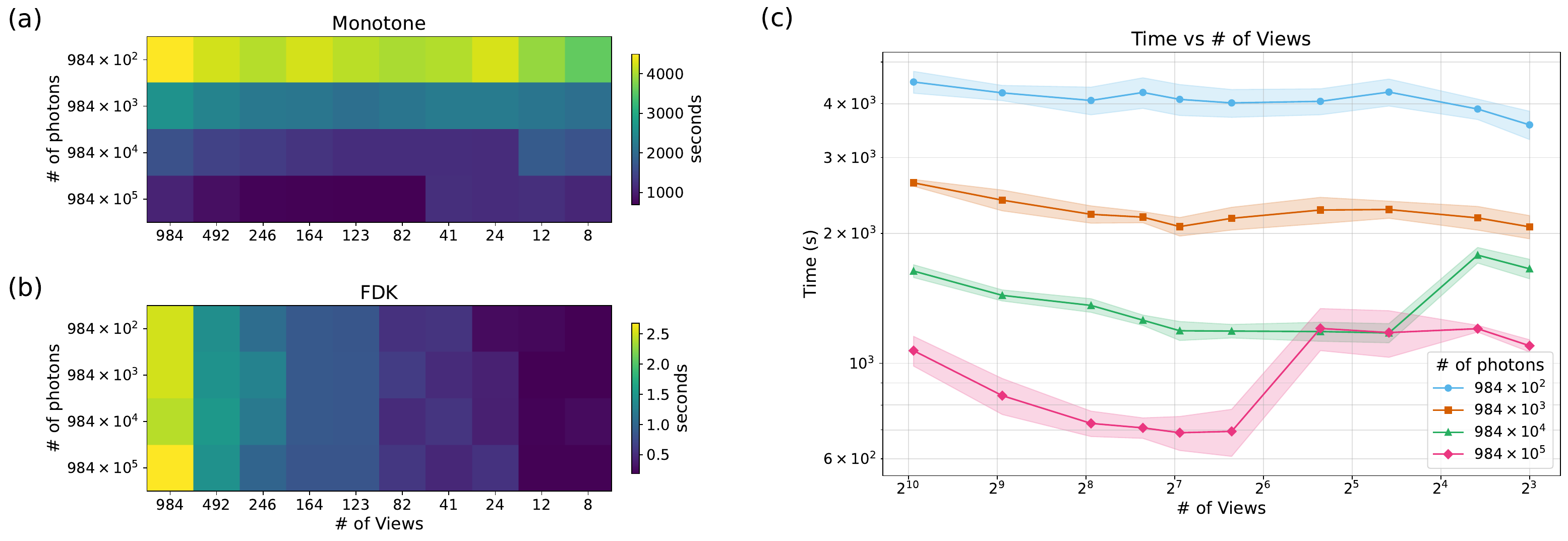}
    \caption{Computation time across dose allocations. \textbf{Left:} Heatmaps of average reconstruction time (s) for \method{} (top) and FBP (bottom) stratified by total photon budget (rows) and number of projections (columns). Each heatmap uses its own color scale, as \method{} is iterative. \textbf{Right:} \method{} reconstruction time vs.\ number of projections for each photon budget; shaded bands show $\pm$1 standard deviation across random seeds.}
    \label{fig:time}
    \end{figure}
    
FBP is a one-step reconstruction algorithm, whereas \method{} is iterative and runs until convergence. Accordingly, we report reconstruction time for each method and each dose allocation setup in \Cref{fig:time}. FBP reconstruction is much faster than \method{} reconstruction, with FBP taking a few seconds and \method{} converging within an hour. The reconstruction time for FBP is primarily a function of the number of projections, with more projections causing longer reconstruction times. In contrast, reconstruction time for \method{} is primarily a function of the total photon budget, with lower-dose reconstructions converging more slowly.

\section{Discussion}

This study demonstrates that a monotone operator based reconstruction algorithm, \method{}, provides consistent improvements over FBP in perfusion PCD CT, for quantification of iodine assuming known background signal. Across a wide range of photon budgets and numbers of projections, \method{} achieves lower RMSE in HU and more favorable noise and SNR characteristics. These improvements were particularly evident in photon limited regimes where FBP reconstructions suffered from severe artifacts and noise amplification. \method{} reconstruction also enables quantitative estimation of iodine concentration, with typical errors below 0.4 mg/ml, for ground truth concentrations up to 2.5 mg/ml and with as small a total photon budget as $984 \times 10^2$. 

The clinical relevance of these findings lies in the potential of PCD CT to deliver high-resolution and spectrally resolved images at reduced dose. Reliable reconstruction under sparse-view or photon-starved conditions is critical for realizing these benefits. Our results suggest that \method{} reconstruction could enable robust quantitative perfusion imaging in settings where conventional FBP fails, potentially supporting dose reduction strategies while preserving diagnostic image quality.

Despite these advantages, several limitations remain. Performance degrades under extreme angular undersampling, highlighting the challenge of severe view sparsity. In addition, the computation time of \method{} scales with the source intensity, requiring longer reconstruction times at lower X-ray doses. Since our evaluation was based on a digital phantom simulation, further validation on physical phantoms and in-vivo data will be required to confirm clinical applicability.

\section{Conclusions}
Future work will focus on accelerating \method{}, extending it to multi-material decompositions, and testing it in experimental and clinical PCD CT acquisitions. Such advances could further establish \method{} reconstruction as a practical and robust alternative to analytical reconstruction in PCD CT.

\subsection*{Acknowledgments} 
AP was supported in part by a Google Research Scholar award and by research awards from Adobe, Amazon, and Mathworks.
SFK was supported in part by the NSF Mathematical Sciences Postdoctoral Research Fellowship under award number 2303178. Any opinions, findings, and conclusions or recommendations expressed in this material are those of the authors and do not necessarily reflect the views of the funding agencies.

\section*{References}
\addcontentsline{toc}{section}{\numberline{}References}
\vspace*{-10mm}

\bibliographystyle{abbrvnat}
\bibliography{reference-template}

\end{document}